\documentclass[10pt,conference]{IEEEtran}
\IEEEoverridecommandlockouts
\usepackage{cite}
\usepackage{amsmath,amssymb,amsfonts}
\usepackage{algorithmic}
\usepackage{graphicx}
\usepackage{textcomp}
\usepackage{xcolor}
\usepackage{xspace}

\usepackage{stfloats}
\usepackage{algorithmic}
\usepackage{graphicx}
\usepackage{textcomp}
\usepackage{xcolor}
\usepackage{xspace}
\usepackage{caption}
\usepackage{graphics}
\usepackage{adjustbox}
\usepackage{multirow}
\usepackage{tabularx}
\usepackage{tikz}
\usepackage{hyperref}
\usepackage{xurl}
\usepackage{pgfplots}
\usepgfplotslibrary{units} 
\pgfplotsset{compat=newest} 
\usepackage{siunitx}
\usepackage{fancyvrb} 
\usepackage{listings}
\definecolor{A1}{HTML}{feeba2}
\definecolor{A2}{HTML}{fed069}
\definecolor{A3}{HTML}{feaa38}
\definecolor{A4}{HTML}{f3801c}
\definecolor{A5}{HTML}{d85a09}
\definecolor{A6}{HTML}{7A2E08}

\definecolor{B1}{HTML}{540804}
\definecolor{B2}{HTML}{81171B}
\definecolor{B3}{HTML}{AD2E24}

\def\BibTeX{{\rm B\kern-.05em{\sc i\kern-.025em b}\kern-.08em
    T\kern-.1667em\lower.7ex\hbox{E}\kern-.125emX}}

\newcommand{\tool}{\textsc{FLowSH}\xspace}

\newcommand{\LONG}[1]
{{}}

\begin{document}
\title{\tool: Detecting Web Attacks by Correlating Anomalous Network Flows and Endpoint Logs}


\author{
    \IEEEauthorblockN{Nuno Durão \ \ \ Rodrigo Branco \ \ \ Vinicius V. Cogo \ \ \ Ib\'{e}ria Medeiros}
    \IEEEauthorblockA{LASIGE, DI, Faculdade de Ci\^encias, Universidade de Lisboa, Portugal \\ nuno.p.durao@gmail.com, fc54457@alunos.ciencias.ulisboa.pt, \{vvcogo, ivmedeiros\}@ciencias.ulisboa.pt}
}


\maketitle

\begin{abstract}

Web applications compose the core of internet services, rendering them prominent targets for audacious attackers. 
Among web attacks, SQL injection (SQLi) remains a persistent, damaging web applications connected to databases.
The widespread adoption of HTTPS has been a critical advancement for online security, ensuring the confidentiality and integrity of communications by encrypting network payloads. 
However, it also renders traditional Network Intrusion Detection Systems (NIDS) less effective, as these typically rely on 
packet inspection of plaintext traffic.
To access encrypted payloads, some NIDS resort to TLS-interception proxies, which leads to data leaks and security risks.
This paper presents \tool, a system for detecting web attacks that is agnostic to whether the traffic is encrypted (HTTPS) or
not (HTTP).
Rather than decrypting packets in transit, \tool correlates flow-level metadata (NetFlows) with endpoint logs, which inherently contain decrypted payloads after reaching the web server. 
The system combines unsupervised machine learning to flag anomalous NetFlows and similarity search over server-side logs using Locality Sensitive Hashing~(LSH) to confirm attacks. 
We implemented and evaluated \tool in a testbed with multiple web scanners and demonstrated its detection effectiveness, detecting 70\% of SQLi attacks missed by ModSecurity, a leading web application firewall. 
\end{abstract}


\section{Introduction}

The recent trend towards remote/hybrid operations propelled a digital transformation in organisations that also entailed a substantial increase in attacks on their services~\cite{CheckPoint:23}. 
Web applications have been massively adopted to compose such a transformation, which leads developers to prioritise functionality at the expense of security, rendering their vulnerable products prominent targets for malicious entities.
The number of registered vulnerabilities has increased 25\% yearly since the pandemic~\cite{cve-date-24}.
For instance, recent SQL injection (SQLi) attacks were reported as the root cause of 4.5M patients having their data stolen from HealthEC~\cite{BleepingC:24}. 
SQLi is ranked as the third most critical vulnerability class by OWASP~\cite{b1}, remaining highly significant both in terms of the volume of reported incidents~\cite{cve-type:24} and their severity upon exploitation.

HTTPS emerged as a pivotal solution to address these security oversights, being extensively utilised to safeguard data communication in web applications through data encryption~\cite{b4}. 
However, despite organisations adopting HTTPS, 
the vulnerable code in the endpoint software (i.e., behind HTTPS) remains exposed and susceptible to malicious attacks.
To increase cybersecurity defence, organisations have employed Network Intrusion Detection Systems (NIDS) to mitigate attacks by inspecting network packet payloads, including web application attacks (or simply web attacks). This inspection process, also known as Deep Packet Inspection (DPI)~\cite{pramod2015sqli,9474271}, although effective, can create \emph{(i) network bottlenecks} because it requires inspecting every packet.
Moreover, the use of HTTPS hinders the detection of such attacks since \emph{(ii) encrypting payloads prevents their examination}.

NIDS typically employ two methods to overcome these obstacles: TLS-interception proxies and Network Flows (\textit{NetFlows}. 
TLS-interception proxies are used when facing encrypted data because they intercept and decrypt network traffic~\cite{Neill:16}, revealing payload data for DPI analysis. 
However, this approach not only \emph{(iii) raises security risks by increasing the attack surface} through the addition of more components with access to the communication parties' cryptographic keys, but it may also aggravate the network bottleneck issue.
Alternatively, NetFlows offer a solution to these issues by enabling packet analysis without individual inspection. 
A NetFlow aggregates a sequence of packets sharing common characteristics into a single flow observed over a period (e.g., $3$ min)~\cite{Hofstede:14}. 
Although NetFlows can provide valuable insights into attack patterns, without weakening transport-layer security, \emph{web attacks may still go undetected}.
%
%
To improve such systems, Machine Learning (ML) has been used in NIDS~\cite{7307098} and NetFlows~\cite{Sacramento:18} for anomaly detection.
These ML-based systems operate over plain-text network traffic, enabling them to identify patterns from attacks they know. 
However, their efficiency is hindered when processing encrypted data.
Therefore, \emph{a system capable of functioning as a NIDS in encrypted and unencrypted traffic scenarios, without weakening transport-layer security, is vital for the security of modern web applications}.

This paper proposes such a desired system -- \tool (\emph{Flow \& LSH NIDS System}), a resilient system for discovering web attacks on network traffic based on similarity search techniques. 
\tool relies upon unsupervised ML to flag anomalous \emph{WebFlows}\footnote{We refer to \emph{WebFlows} as specialised NetFlows carrying web application communication, differentiating them from the general NetFlow definition.} and similarity search, powered by \emph{Locality Sensitive Hashing} (LSH), that acts as DPI to identify web attacks in this traffic.
The system overcomes the encrypted data challenge by retrieving the traffic information from the web server logs, without weakening transport-layer security.
More specifically, \tool resorts to centroid-based clustering models to swiftly identify clusters with suspicious WebFlows and pinpoint their traffic for deep analysis.
The payload's data is then queried on an LSH database to confirm its web attack status.
LSH~\cite{indyk1998approximate,Gionis99LSH} is an Approximate Nearest Neighbour (ANN) solution that employs hashing techniques to group data points into buckets according to their similarity.
Buckets work as LSH databases that efficiently respond to data point queries with a small set of candidate points.

Our current implementation focuses on detecting SQLi attacks in web applications, dealing with HTTP(S) protocols.
We conducted experimental evaluations for attesting \tool's robustness and a comparative evaluation with ModSecurity, a prominent web application firewall, where \tool detected 70\% of the attacks that ModSecurity missed.

The main contributions of the paper are:
(1) a reliable system for detecting web attacks over network traffic, including HTTPS~(\S\ref{sec:approach});
(2) a novel approach based on unsupervised ML for flagging anomalous traffic and LSH techniques for confirming web attacks on such traffic (\S\ref{sec:approach});
(3) \tool, an implementation of the approach (\S\ref{sec:implementation}) and its evaluation for SQLi attacks (\S\ref{sec:evaluation}).

\LONG{
\subsection{Challenges}
There are three main challenges that we want to address in this paper.

The first is understanding how to capture suspicious HTTPS Network Traffic related to Web Application Attacks. The security provided by HTTPS protocol prevents one from having access to the plain-text payload during network traffic analysis. Decrypting HTTPS packets is a way to have access to such content. The decryption can be achieved using a proxy server that captures all the traffic, decrypts accesses and re-encrypts it. This approach uses DPI, or Deep Packet Inspection techniques, to inspect their content. The problem with this approach is that it tends to create a bottleneck in the network and decreases the overall security. The bottleneck is caused by decrypting data and evaluating it in real-time, which is a processor-intensive task that many hardware-based security devices need help handling.

Moreover, the fact that there is an additional component where the payload resides decrypted between the client and the final server increases the attack surface, decreasing the encryption strength and threatening the overall data security of HTTPS traffic. A possible solution for this challenge would be using NetFlows. NetFlow is a protocol for monitoring network flows and collecting IP traffic statistics. Using a NetFlow monitoring system enables more efficient and effective monitoring and analysis of these flow records about network traffic. More specifically, we use WebFlows throughout this paper to clarify when it deals with web network traffic.

The second challenge addressed in this paper is the Web Application Attacks Constraints. In terms of web attack detection, some attacks are detectable through the observations of a WebFlow. That is the case of brute force attacks, for example, HTTP(S) dictionary attacks. On the other spectrum, there are attacks where we need to analyse the payload content to detect the attack. SQLi and XSS are examples of these types of attacks. However, we cannot directly identify these in a WebFlow because we cannot access the payload.

The third and last challenge is understanding how an LSH database built with inputs from specific scanners and fuzzers can identify other inputs outside the ones created by these tools. In this paper, we will explain how we can detect malicious inputs using a Similarity Search Engine such as the LSH and how different inputs created by different fuzzers and scanners can influence the recall of our system.
} 

\section{Concepts and Related Work}
\label{sec:rw}

This section introduces the main background concepts and related works essential for understanding our proposal.



\LONG{
To study the Network Traffic, knowing about TPC/IP protocol and the difference between HTTP and HTTPS protocol is necessary.

\subsubsection{TCP/IP Protocol}
Transmission Control Protocol/Internet Protocol model functionality is based on four layers, each with its own set of protocols, and each one communicates with its adjacents. The layers are organised in this way:

\begin{itemize}
  \item \textit{Application Layer}. It contains higher-level protocols such as File Transfer Protocol (FTP), Simple Mail Transfer Protocol (SMTP) and Hyper Text Transfer Protocol (HTTP). 
  \item \textit{Host-to-Host Layer}. TCP and UDP are the two protocols found in this layer. Used for protecting upper-layer applications from data complexity.
  \item \textit{Internet Layer}. In this layer, Internet Protocol implements a set of rules for routing and addressing datagrams. A datagram contains a header, source and destination address, and data.
  \item \textit{Network Interface}. This layer characterises how data is physically transported via the network.
\end{itemize}
}

\LONG{
\subsubsection{HTTP and HTTPS}

The HTTP protocol~\cite{rfc2616} is a client-server or request-response protocol. A user-agent usually makes the requests and provides an answer that sends a response to the client. The protocol is generic and stateless and works in the TCP/IP's application layer.
%
HTTPS protocol~\cite{rfc2818} is a version of HTTP that runs over TLS~\cite{rfc2246} and is responsible for the encryption and authentication of the data exchanged by the application layer. HTTPS protocol means that HTTP is enclosed in a cryptographic scheme to provide security to the communication over the network, enforcing confidentiality, integrity and availability (CIA). 
%
The use of HTTPS allows the protection against man-in-the-middle attacks, eavesdropping and tampering. According to W3Techs \cite{b4}, 85\% of all websites use HTTPS, and the tendency is to rise. 

For example, Pramod~\emph{et al.}~\cite{pramod2015sqli} used DPI to detect SQLi attacks but just on HTTP because on HTTPS the payload is encrypted. 
Additionally, there are some direct and noticeable protocol-related performance costs related to HTTPS, such as the increasing latency. Naylor~\emph{et al.}~\cite{HTTPS_costs} identified some indirect consequences in in-network services like loss of caching and on some services such as parental control or virus scanning. 
One major result from their published study was that although HTTPS messages are bigger in size, the vital concern is with the TLS handshake because in 50\% of TLS communication they study, the hand-shake represents more than 42\% of the total data exchanged.
}

\vspace{2mm}
\noindent
\textbf{Network Flows.}
\label{ssec:RW_netflow}
NetFlows~\cite{Hofstede:14} are aggregates of network traffic over a time interval, collected, processed, and exported by monitoring tools.
Packets within a NetFlow share header attributes (e.g., destination IP and port and application-level headers) and derived characteristics like length and processing details.
Cisco's NetFlow v9~\cite{netflow9} defines the main standard, while IPFIX has evolved it with features such as transport options, variable-length strings, enhanced security, and template withdrawal~\cite{itfnews}, which have made them widely adopted for flow collection and analysis~\cite{7509765}.

\vspace{2mm}
\noindent
\textbf{SQLi Attacks.}
%
SQL Injection (SQLi) is an attack on web applications where crafted SQL code is injected into queries, changing the query's logic and causing unintended data exposure upon execution~\cite{b1}.
%
Listing~\ref{lst:sqli} shows a SQLi vulnerable PHP code that authenticates users based on submitted credentials (lines 1--2).
If an attacker inputs \Verb+' OR 1=1;-- + as the username, the resulting query (in line 3), when executed (line 4), retrieves all user records from the database.

\begin{minipage}{\columnwidth}
\begin{lstlisting}%
[language=PHP,showstringspaces=false, xleftmargin=-3mm, xrightmargin=4mm, caption={\small A PHP code with an SQLi vulnerability.}, label={lst:sqli}]
 $u = $_POST['username'];
 $p = $_POST['password'];
 $q = "SELECT * FROM users WHERE user='".$u."' AND pass='".$p."'";
 $result = mysqli_query($conn, $q);
\end{lstlisting}
\end{minipage}

\vspace{2mm}
\noindent
\textbf{Similarity Search with LSH.}
\label{ssec:LSH}
Locality-Sensitive Hashing is a widely used \textit{Approximate Nearest Neighbour} (ANN) approach for \textit{Similarity search} that retrieves items based on resemblance~\cite{indyk1998approximate,Gionis99LSH} and with scalable capacity.
Through ANN, LSH hashes similar items into the same buckets with high probability.
Queries in LSH return small sets of candidates that likely resemble the query.

To use LSH, objects are first decomposed into \emph{$k$-shingles}, a sequence with a fixed size of contiguous tokens.
These may represent any object, such as bytes, strings, or pixels, depending on context.
For example, the string~\verb|abcdabd| yields the set \verb|{ab,bc,cd,da,bd}| for $k=2$~\cite{leskovec:20}.
The similarity between two objects is typically computed using distance metrics like Jaccard, where similarity is measured by $|S \cap T| / |S \cup T|$~\cite{jaccard-similarity}.

Direct shingle comparison is inefficient for large datasets, so LSH hashes shingles into integer arrays that represent objects.
Each array is split into subsets, based on a similarity threshold, that serve as keys to multimaps storing candidate buckets.
These LSH hash families preserve similarity, with MinHash~\cite{Broder97Minhash} being the most used, aligned with Jaccard, and use configurable permutations to produce hash values that reflect set similarity through minimal differences~\cite{indyk1998approximate}. 


\vspace{2mm}
\noindent
\textbf{Network Intrusion Detection Systems.}
%
Deep Packet Inspection (DPI) and NetFlows are the two main approaches to detecting anomalous traffic, the former being the most employed by Network Intrusion Detection Systems (NIDS).

There are diverse commercial products and research approaches to DPI. For instance, Rescio~\emph{et al.}~\cite{9474271} presented a benchmark of the major commercial products and listed several studies.
Murgai~\emph{et al.}~\cite{Murgai:24} inspect the network traffic using NIDS Snort rules as features for ML models to detect anomalies.
Pramod~\emph{et al.}~\cite{pramod2015sqli} used DPI to detect SQLi attacks. However, these works inspect clear data, like HTTP.

With the rise of HTTPS encrypted communication, new network security monitoring and analysis issues have arisen.
One of the solutions often presented is the use of a proxy server. 
Jarmoc~\cite{jarmoc} discusses how a proxy server works and its associated risks. 
Interception proxies inject themselves into the traffic flow and terminate the clients' requests to inspect conversations via SSL in plain-text. The purpose is to provide access to the plain-text session data to be analysed by DPI, but it can simultaneously create security issues.   
Branco~\emph{et al.}~\cite{branco24enase} proposed detecting web attacks through log-based analysis using NLP and supervised learning, but do not evaluate the NetFlow-based anomaly detection they describe.
Contrary to proxies, Sherry~\emph{et al.}~\cite{blindbox} proposed BlindBox, a DPI solution to TLS with privacy models preventing this middlebox from accessing unidentified traffic as an attack by the rule generator. However, it replaces HTTPS with a new end-to-end encryption protocol, which is infeasible for many other systems.

There is much work done resorting NetFlow to detect web attacks, but not specific web application attacks, like SQLi.
Some works discussed the use of NetFlows to detect DoS and DDoS~\cite{5455789, Sacramento:18} and Network worm virus traffic~\cite{5455789}.
Other works~\cite{300408947,8546856} studied the use of NetFlow to detect SSH Brute force attacks in both HTTP and HTTPS.
Some of these works resort to ML to detect such anomalies~\cite{Sacramento:18}. 
The sparse existing work addressing HTTPS~\cite{7140395,300408947} involves patterns of HTTPS dictionary attacks in NetFlow. 
However, their approach analyses SSH brute force attacks, not web application attacks such as SQLi.

Unlike these works, we propose an approach involving WebFlows and DPI to detect web application attacks on HTTP(S) traffic. For WebFlows, the approach resorts to unsupervised ML, whereas for DPI, a similarity search engine based on LSH is used. To the best of our knowledge, such an engine has never been used in DPI, nor in combination with WebFlows (even NetFlows).

\LONG{

Our goal is to create a NIDS to detect web attacks coming on HTTP and HTTPS network traffic by analysing WebFlows and packet payloads using, respectively, unsupervised ML and LSH-based DPI, without weakening transport-layer security. 
Before introducing \tool, we present the challenges we faced in building it.

\subsection{Challenges}

\noindent\emph{1) How to capture suspicious HTTP(S) network traffic related to web application attacks?}
As we intend to create a NIDS integrating DPI, we must inspect packet payloads in plain-text. This raises two issues:
(i) DPI introduces performance bottlenecks by inspecting every packet, which is unsuitable for high-speed networks;
(ii) HTTPS traffic hides the payloads, requiring decryption, which compromises security and adds complexity.

NetFlows offer a lightweight alternative by aggregating bidirectional traffic based on transport and network layer features. 
WebFlows, a subset of NetFlows specific to web traffic, can capture HTTP(S) activity without deep inspection. 
However, identifying suspicious flows remains a challenge. 

Web scanners, commonly used to exploit application vulnerabilities, generate high-volume traffic. 
We propose using unsupervised ML to cluster WebFlows and flag outliers, such as scanner traffic, as anomalous.

\noindent\emph{2) How to address web application-layer attack detection?}
Some attacks (e.g., brute force, dictionary attacks) can be inferred from flow metadata, but others (like SQLi) require payload inspection.
To detect these web attacks, while preserving transport-layer confidentiality, we leverage web server logs, which inherently contain decrypted payloads at the endpoint.
By correlating these logs with previously identified anomalous WebFlows, we inspect only relevant traffic, mitigating both DPI overhead and HTTPS decryption issues.

\noindent\emph{3) How can DPI timely detect attacks across diverse scanners?}

Scanners often vary their payloads using different encodings or keywords. Traditional pattern-matching DPI must maintain extensive signature databases and perform costly lookups.

Our novel DPI method replaces them with a Similarity Search Engine (SSE) based on LSH, enabling fast, approximate matching over large payload datasets. 
We build LSH databases from scanner-generated attacks and query them using tokens extracted from suspicious flows.
\tool uses decrypted payloads available in server logs, avoiding the need for TLS key access and preserving HTTPS security. 

\subsection{Approach Overview}
\label{ssec:overview}

Figure~\ref{fig:pipeline} shows the \tool NIDS architecture we propose, which is divided into three main phases: \emph{Suspicious Traffic Detection}, \emph{Web Attack Identification}, and \emph{Alarm Notification \& Continuous Improvement}.

The first phase, the \emph{Suspicious Traffic Detection}, identifies anomalous HTTP(S) WebFlows associated with Clients activity and the respective payloads through a five-step pipeline performed by the \emph{WebFlowProc} and \emph{LogExtractor} entities. 
The \textit{WebFlowProc} captures the traffic generated from the interactions between Client-WebApp entities. A Client is any entity that can interact with a WebApp, e.g., a normal user, an attacker, or a scanner. It then separates this traffic into WebFlows and groups them into clusters using an unsupervised ML algorithm. The clusters that contain only a few WebFlows are particularly interesting as they may contain abnormal traffic, such as attacks generated by scanners. 
These WebFlows are marked for further analysis by the second entity.
At this point, the amount of data processed was significantly reduced, leaving only a few WebFlows that must be inspected.
The \textit{LogExtractor} filters the WebLogger's logs to find the payloads of these WebFlows and retrieves: the data that was sent by the Client (i.e., the malicious inputs) to the WebApp, the URL, and the parameters involved in the attack (e.g., the fields of WebApp's forms where the data was injected).

The \emph{Web Attack Identification} phase verifies if the suspicious WebFlows carry known web attacks similar to those contained in the LSH database (the \textit{Tokens DB} in Figure~\ref{fig:pipeline}).
To do so, the \textit{SimilarSearchEngine} (SSE) entity performs two steps. 
First, the data retrieved by the LogExtractor entity is tokenised following the shingling process used in the LSH-database construction, enabling LSH-queries on such a database. 
Next, LSH-queries are composed of these tokens and executed on the database for a defined threshold, where all database tokens with a similarity level greater than or equal to the query's tokens will be returned. 

\begin{figure*}[t]
    \includegraphics[width=\textwidth]{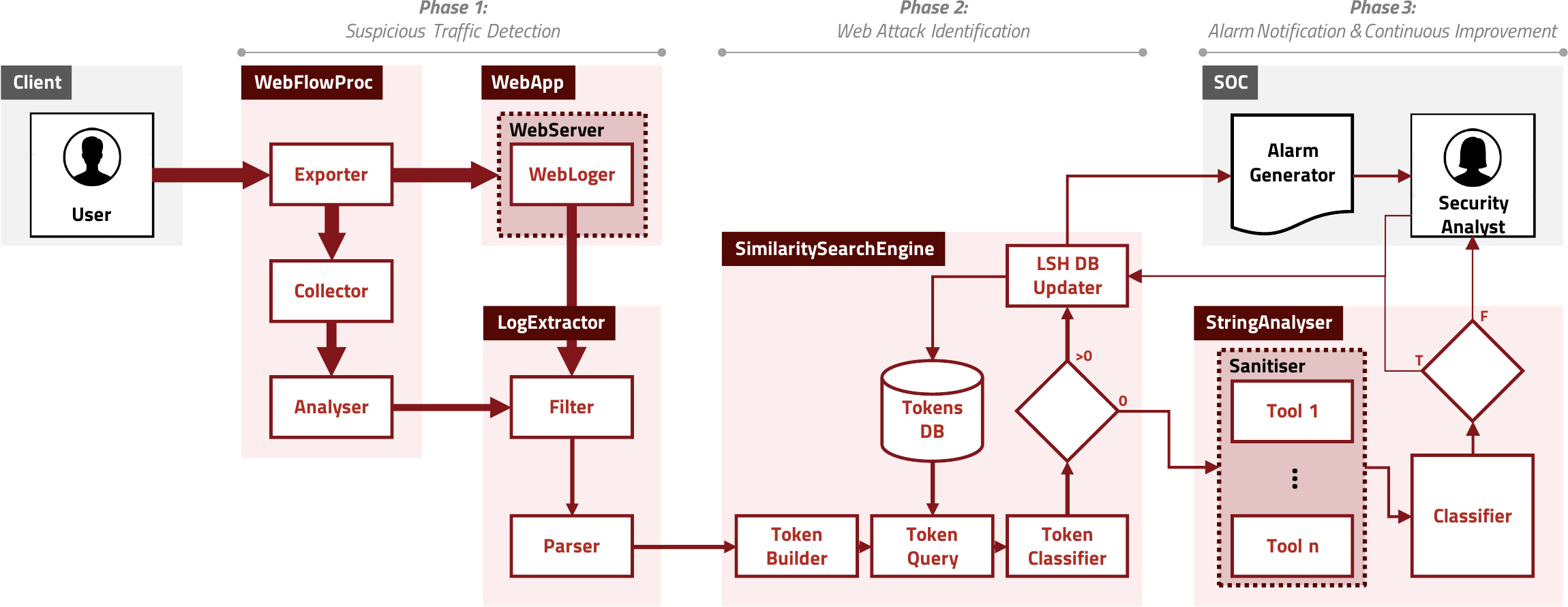}
    \caption{\small Architecture of the proposed \tool solution to detect web attacks over HTTP(S) network traffic. Arrow thickness indicates the relative amount of data flowing among components.}
    \label{fig:pipeline}
    \vspace{-5mm}
\end{figure*}

Lastly, the \emph{Alarm Notification \& Continuous Improvement} phase analyses the query results and, based on that, notifies the Security Operation Center (\textit{SOC}) entity when attacks are confirmed and updates the Tokens DB when new attacks are found. In the case of undecidability of the presence of an attack (e.g., attacks that were unseen in the Tokens DB), the query tokens are checked with a set of string analysers, which perform string sanitisation and validation tasks used in web applications to invalidate web attacks. The database is updated if this verification is positive; otherwise, the security analyst determines whether an attack has occurred or not, updating the database accordingly.
} 

\section{\tool Approach \& Design Insights}
\label{sec:approach}

Our approach aims to enhance the defence capabilities of organisations against web attacks on web applications that support, for example, the services they provide and use, and help SOC teams in preventing malicious actions against organisations. To conduct that, we propose \tool to detect web attacks coming on HTTP(S) network traffic by analysing WebFlows and packet payloads using, respectively, unsupervised ML and LSH-based DPI, without weakening transport-layer security. It begins by gathering the interactions between Client-WebApp entities and then inspects the packet payloads it finds as suspicious. Clients are entities that interact with a WebApp, e.g., a normal user, an attacker, or a web scanner.

Resorting to unsupervised ML and LSH-based DPI, \tool handles the aforementioned problems $(i)$ -- $(iii)$ by resorting to WebFlow analysis, which can flag possible anomalous traffic without the need to access and inspect the application data layer and consequently only demands further inspection of this traffic, i.e., their payloads. For payload analysis, it first accesses web server logs, where payloads are registered in plaintext (after the web server receives the traffic and decrypts it), and then it searches LSH databases for similar payloads, thus detecting attacks if a match is found. Identified attacks generate alarms to SOC, and unseen attacks identified by SOC are added to LSH databases. The approach acts continuously in a loop, updating \tool with attack knowledge and database enrichment.
Figure~\ref{fig:pipeline} shows the \tool NIDS architecture we propose, which is divided into three main phases and summarised as follows:

\begin{enumerate}
    \item \emph{Suspicious Traffic Detection:} identifies anomalous HTTP(S) WebFlows associated with Clients activity and extracts the respective payloads from web logs. At this point, data was significantly reduced, leaving only a few WebFlows that must be inspected.

    \item \emph{Web Attack Identification:} verifies if the suspicious WebFlows' payloads carry known web attacks similar to those contained in the LSH database (the \textit{Tokens DB} in Figure~\ref{fig:pipeline}). To do so, LSH queries are composed from payloads and submitted to the Tokens DB.
    
    \item \emph{Alarm Notification \& Continuous Improvement:} analyses the query results and, based on that, notifies the SOC entity when attacks are confirmed and updates the Tokens DB when new attacks are found.
\end{enumerate}

The next sections explain the phases in detail.

\begin{figure*}[t]
    \includegraphics[width=\textwidth]{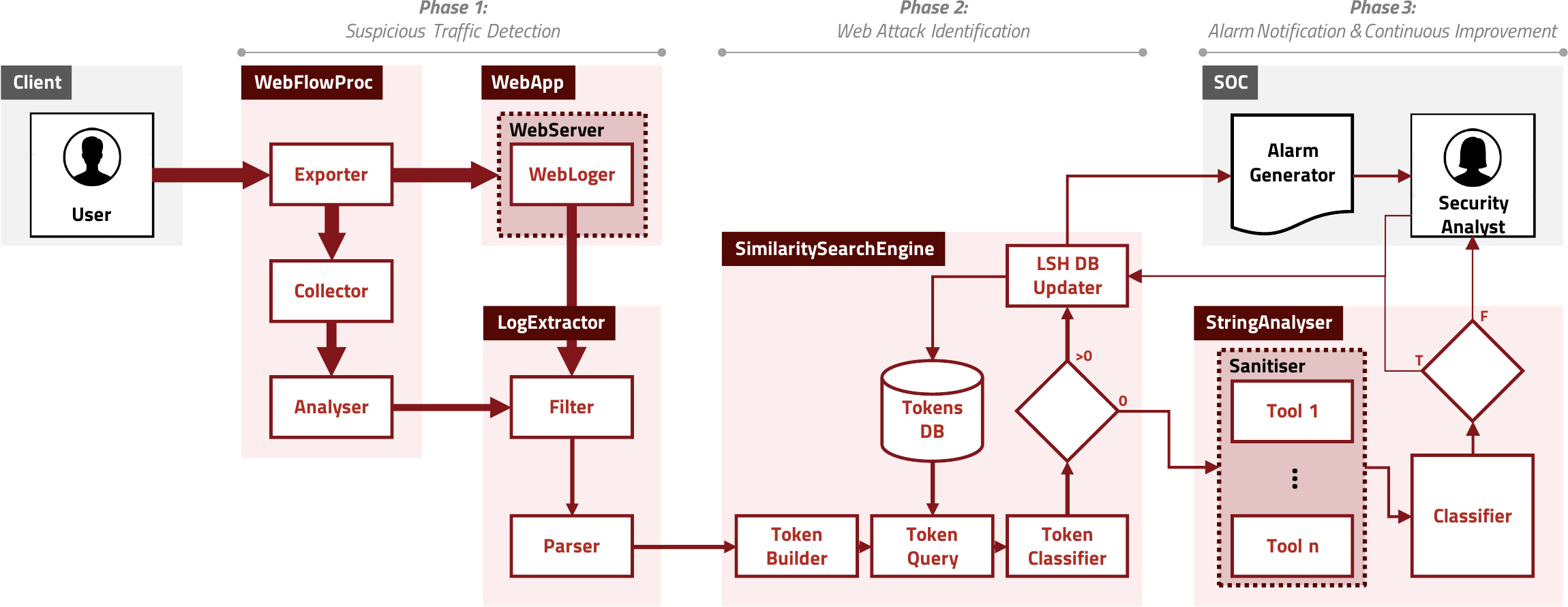}
    \caption{\small Architecture of the proposed \tool solution to detect web attacks over HTTP(S) network traffic. Arrow thickness indicates the relative amount of data flowing among components.}
    \label{fig:pipeline}
    \vspace{-5mm}
\end{figure*}

\subsection{Phase 1: Suspicious Traffic Detection}
\label{ssec:phase1}

This first phase detects anomalous traffic, such as those originated from scanners, through WebFlows clustering performed by unsupervised ML, and then extracts their respective payloads. 
These actions are conducted by the \emph{WebFlow Processor} and \emph{Log Extractor} entities through a five-stage pipeline and interactions with the other two entities -- \emph{Client} and \emph{WebApp}.



\subsubsection{WebFlow Processor}

\paragraph{Exporter and Collector}
%
HTTP(S) network traffic is captured in the form of PCAP and converted into a WebFlow. A WebFlow condenses bi-directional traffic of Client-WebApp in a single flow, occurring in a time period (e.g., $3$ min). The structure of a WebFlow is the same as a NetFlow, i.e., a set of features that are present in the network and transport layers of TCP/IP protocol, such as source and destination IP and ports, and others that are calculated, for instance, the number of packets and bytes exchanged in that period of time. 
However, as some of these features are expressed as text (e.g., IP addresses, protocol, and protocol flags) and ML algorithms require numeric data to process them, these features are converted to numbers.

\paragraph{Analyser}
%
WebFlows are processed by the K-Means clustering algorithm, which aggregates WebFlows with similar characteristics into the same group, thus forming clusters. Smaller clusters are considered suspicious, as by the norm, they deviate from normal traffic patterns.



Clustering is commonly used in anomaly detection, where clusters 
with few instances are considered anomalies (i.e., outliers) since their instance values deviate from the normal and expected values. 
In real-world scenarios, normal traffic is expectedly larger than the traffic generated by scanners, and, as the traffic from these considerably differs from the expected traffic, the resulting clusters will group traffic with similar features, like source and volume, two of the significant features that identify a scanner. 

K-Means~\cite{Witten:25} aggregates the WebFlows into $K$ clusters.
However, as it requires a \emph{priori} the number $K$ clusters to create, we resort to the \textit{Elbow} method~\cite{Witten:25} to estimate it.
%
Additionally, the algorithm requires that the scale of all WebFlow features be standardised as it relies upon the distance between points, and WebFlows have features with different unit measures. To accomplish such a requirement, we use \textit{StandardScaler} and \textit{MinMaxScaler}~\cite{scikit-learn} to standardise the distribution to a zero mean and a one standard deviation. 
Thus, each feature is scaled by subtracting the mean and dividing by the standard deviation. 



The result of the \emph{Analyser} step is a set of $K$ clusters, where those with few WebFlows are selected for payload inspection.
Note that the quantity of payloads to be analysed will be reduced again since this step has the ability to segregate the traffic produced by the scanners.
%

\subsubsection{Log Extractor}

To analyse the payload, we need to access the original one without additional resources that could introduce weaknesses to \tool, such as intermediate proxies to decrypt HTTPS traffic. 
To accomplish this, we resort to the Web Server logs and must capture the payloads when traffic arrives at the web server that stores the web application. When the web server receives the traffic, it decrypts it (in the case of HTTPS) so the web application can process the requests and log the interactions derived from them. 
Also, we must ensure that all the traffic is logged by the server, as we want to analyse the content of all types of payloads. 
%
%
However, analysing these logs is not straightforward, and we identified some problems:
$(i)$ POST requests are not logged by the web server. For example, the Apache server, by default, logs the content from GET requests but not from POST requests. These types of requests are the ones that contain user input or scanners' data, and so are the most interesting to access;
$(ii)$ Logs register all communication steps, which hardens the task of discovering attacks in a fast and efficient way;
$(iii)$ The log format is not the best for conducting such inspections.
To cope with these issues, our approach includes the \emph{mod\_dumpio} module to register and capture POST requests, 
and a \textit{Filter} and a \textit{Parser} to, respectively, obtain only those logs containing user data from GET/POST requests and transform them into a format that is easier to process.

\paragraph{Filter}
This step retrieves the subset of logs corresponding to the identified anomalous WebFlows.
To do so, it uses as matching keys the \textit{datetime}, \textit{source IP} and \textit{destination IP} since WebFlows and logs have these keys in common.
However, the \textit{datetime} is not the same in both structures because the traffic of a WebFlow is captured a moment before it arrives at the web server, but it is close enough.
The filter considers this deviation through a threshold (e.g., 5 secs) to ensure it gets all logs related to the identified traffic.
It also discards logs that do not have payloads (e.g., connection establishment, SSL handshake), keeping only logs with GET/POST requests.

\paragraph{Parser}
The log data comes in chunks, meaning that the payload is split into fragments, which makes its analysis difficult. Therefore, chunks need to be processed to obtain a steady stream of logs containing all the requests and their payloads.
To do so, the \textit{Parser} step parses the chunks, identifies the GET/POST requests, joins all the parts of each request and rebuilds the payloads.

\subsection{Phase 2: Web Attack Identification}
\label{ssec:phase2}

This phase is performed by the \emph{SimilaritySearchEngine} entity, which involves a three-step pipeline and a \emph{Tokens DB} (see Fig.~\ref{fig:pipeline}) that inspects the payloads of the anomalous WebFlows against an LSH database (\emph{Tokens DB}) of attacks to confirm the presence of web attacks (e.g., SQLi). 
We propose a novel DPI method based on LSH 
and an SSE that integrates this DPI. Unlike traditional DPIs, SSE avoids bottlenecks and performs searches in a reasonable time.
Since LSH operation relies on similarity measures, the LSH-DPI resorts to the MinHash (min-wise independent permutations) algorithm with the Jaccard similarity measure to determine which of the Token DB's instances best match the payload we are searching.
%

\paragraph{Token Builder}
This step standardises the anomalous WebFlows' payloads into tokens.
It removes everything but the meaningful values from each payload. 
Then, each set of values is transformed into a unique token. To do so, each value is transformed into \textit{k-shingles} and encoded into the LSH object -- a hash. 
However, to transform payloads into LSH objects, we must consider the concepts behind LSH (see Section~\ref{ssec:LSH}) and a set of assumptions:
$i) $Although in the SSE the objects to search are payloads, the representation of an LSH hash is a sequence of integers, and because of that, the input data is standardised in tokens. 
By creating tokens, payloads with the same short pieces will have many common elements. 
$ii)$ So, for the shingling process, we have to decide which part of the payload we want to process (i.e., fields, values, or both) to be later searchable and what will be the length of the shingles (\textit{k-shingle}). 
For example, in the key-value pair $password=ZAP$, which part is of interest to represent in an LSH object: the pair or just the value? 
The k-shingle transforms the token into a continuous sequence of unique $k$ items; the value of $k$ should be chosen so that the likelihood of any given shingle occurring in any given payload is minimal.
$iii)$ By segregating the payloads in shingles, payloads with short sections in common will have higher similarity. 

Our initial analysis indicates that only values should be represented, and shingles with length $k=3$ result in better searches. 
%
An example of this processing is depicted in Listing~\ref{lst:completeExamplePayload}, where the values of the payload at the top are transformed into the token at the bottom.

\begin{lstlisting}[breaklines, numbers=none, caption={\small A payload example (top) and its resulting token (bottom).},label={lst:completeExamplePayload}, escapeinside={(*@}{@*)}]
(*@\textbf{Pay:}@*) login=ZAP&email=ZAP&password=ZAP&password_conf=ZAP&secret=ZAP+OR+1\%3D1+--+&mail_activation=&action=create

(*@\textbf{Tok:}@*) [ZAP,AP+,P+O,+OR,OR+,R+1,+1\,1\%,\%3,%3D,3D1,D1+,1+-,+--, --+,cre,rea,eat,ate]
\end{lstlisting}



\paragraph{Token Query}


This step searches for objects in \textit{Token DB} with a similarity to the query of at least a certain threshold.
\tool's LSH includes three key tunable parameters: the hash function, which generates similar hashes for similar items; the similarity threshold determines the required similarity level between objects; and the number of permutations in the LSH, where more permutations reduce error rates.
%
For example, taking the token resulting from Listing~\ref{lst:completeExamplePayload} as the token we want to search for and setting a threshold of $70$\%, this step searches the \textit{Token DB} for tokens with a similarity level of at least $70$\% compared to that token. 
This step interacts with a large database that can be customised on these key parameters.
\paragraph{Token Classifier}

In this step, it is determined whether the queried token (the payload in analysis) is a web attack or not (i.e., the token is benign) or it needs further analysis (explained in Section~\ref{ssec:phase3}). 
The decision is based on the quantity of tokens returned by the Token Query. If it returns zero tokens, the queried token is forwarded for deep analysis because the token can be benign or an attack that does not belong to the Token DB.
Otherwise, depending on the level of similarity between the returned and queried tokens, the queried tokens should be further inspected, or an alert should be generated and sent to the SOC analyst. 
These cases are detailed in the next phase. 



\paragraph{Tokens DB}
It is an LSH database of traffic produced by four web scanners (ZAP~\cite{b7}, Wapiti~\cite{b8}, w4af~\cite{b9}, and Burp~\cite{burp}), where its tokens
are created by the same procedure explained in the Token Builder step.
After running the scanners, the logs from the web server were retrieved, processed to keep only useful information, and the tokens were created from this information and inserted into the database. 
The database is configurable for values of threshold and permutation, allowing the construction of multiple databases to accommodate different targeted and tailored configurations, which in turn permits a reduction in error rates.




\subsection{Phase 3: Alarm Notification \& Continuous Improvement}
\label{ssec:phase3}

The last phase of the \tool manages alarm notifications to the SOC analyst when web attacks are detected, and Token DB updates when new attacks are identified but not yet in the database. 
This continuous improvement, in addition to the database updating, will reduce SOC interaction since, over time, the database will be enriched with knowledge of new attacks.
These tasks are carried out by the \emph{String Analyser} entity and the \emph{LSH DB Updater} step, both of which interact with the \emph{SOC} entity, and their execution is based on the decision of the \emph{Token Classifier}.
%
The \emph{Token Classifier}, upon receiving the tokens from a query, triggers the \emph{String Analyser} when the number of tokens it receives is zero, or the \emph{LSH DB Updater} otherwise.

\subsubsection{String Analyser}

If the query returns no token, either $(i)$ it is an attack but does not belong to the Token DB; $(ii)$ it is benign.
To avoid uncertainty, this module deeply inspects the present token  
by using different string analysers and decides based on the obtained results.
To conduct this inspection and decision, the module contains a \textit{Sanitiser} and a \textit{Classifier} that perform these actions, respectively. 

\textit{Sanitiser} executes a set of tools for looking for dangerous metacharacters in the payloads. These tools execute sanitisation functions on data, which depend on the user's needs regarding the attack classes that it intends to be supported by \tool. 
For example, if the user intends to detect XSS and SQLi attacks, Sanitiser will include tools for data sanitisation related to these attacks. 
\textit{Classifier}, in turn, verifies if the original payload differs from the sanitised one, and so it is considered an attack (scenario $(i)$) and sent to the \textit{LSH DB Updater} to be inserted in the \textit{Token DB}; otherwise, it is benign (scenario $(ii)$) and sent to the SOC analyst for verification. 
If the analyst considers the payload an attack, the associated token is added to the database by the \textit{LSH DB Updater}.



\subsubsection{LSH DB Updater}

If the query returns tokens, an attack was detected and the queried token either $(i)$ is equal to a returned token or $(ii)$ is similar to one of them.
In $(i)$, an alarm is generated and sent to SOC, notifying that a web attack was detected. No other action is taken as the token already exists in the database. 
In $(ii)$, a confirmation is conducted to decide if the queried token is an effective attack and added to the database, and an alarm can be  provided. 
For cases where the level of similarity is higher than a very high defined threshold (e.g., $95$\%) or it is very close to the queried token, they are also considered attacks. An alarm is then generated and sent to SOC. Additionally, the Tokens DB is updated, and SOC is notified accordingly, in cases where tokens in the database differ from the present token but exhibit similarity within certain thresholds.  
Lastly, for cases where the similarity level is near the threshold, the \textit{exact Jaccard similarity} between tokens is calculated and deemed an attack if it exceeds a given threshold. In such a case, an alarm is raised, and the token is added to the database.
Note that LSH is a probabilistic algorithm, meaning that the Jaccard similarity measure it provides is a probability. The \textit{exact Jaccard similarity}, in contrast, is deterministic, and so is the effective Jaccard similarity.  

%


\section{Implementation}
\label{sec:implementation}

\tool implements its architecture through four main modules: \emph{Network Traffic Capture and Processing}, \emph{Log Extraction and Parsing}, \emph{Similarity Search Engine}, and \emph{String Analyser and Sanitiser}. The first two modules correspond to Phase 1, while the other two modules correspond to the other two phases, respectively. 
The system is developed in Python v3.1, which integrates a set of scripts developed from scratch to perform tasks of parsing, correlation, and integration of all modules, and  existing tools to detect SQLi attacks.

\tool uses the Wireshark~\cite{b11} and \texttt{tcpdump}\cite{b12} to capture HTTP(S) traffic and produce their PCAPs, the SiLK framework\cite{b13} to convert these into WebFlows, and K-Means (from scikit-learn~\cite{scikit}) to cluster WebFlows. 
The Apache web server allocates the web applications to be monitored by \tool, and it was configured with the \texttt{mod\_dumpio} module to log POST requests.
%
SSE integrates the LSH DPI and database resorted on \texttt{datasketch}~library~\cite{b15} and configured with MinHash (3-gram tokens), Jaccard similarity, and tunable parameters: permutation $p$ and threshold $t$. 
The \textit{Sanitiser} module uses MySQLdb~\cite{b17}, a Python API for MySQL, to sanitise payloads via \textit{mysql\_real\_escape\_string\_quote}.

\section{Experimental Evaluation}
\label{sec:evaluation}

This section evaluates \tool on the following aspects:
$Q1.$ How effectively does \tool detect anomalous traffic potentially containing SQLi attacks?
$Q2.$ Can one LSH database help identify SQLi attacks generated by various web scanners?
$Q3.$ Does \tool accurately confirm the presence of attack payloads in the detected anomalous WebFlows?

\subsection{Evaluation Setup}
\label{ssec:deploy}

This section describes the experimental testbed used and the LSH-token databases constructed to evaluate \tool. It also presents the process used to confirm the results of the queries and determine the recall and precision metrics.

\paragraph{Experimental Testbed Setup}
%
We used a controlled network environment, with the precision and reliability of the results ensured by our traffic generation. 
Its setup comprises web scanners and vulnerable web applications deployed in three virtual machines (VMs). 
One VM serves as the web server, while the other two VMs act as clients. The server and one client operated on Debian-based distributions with $2$ vCPUs and $4$GB of RAM.
The other client ran on an Ubuntu-based distribution with $4$ vCPUs and $8$GB of RAM.
VMs were hosted on a Windows 64-bit node with one AMD Ryzen 5 4600H CPU ($6$ cores at 3GHz, $12$ threads) and $16$GB of RAM.

The server VM hosts the Apache v2.4.38 equipped with the \texttt{mod\_dumpio} add-on and allocates the BWAPP~\cite{b10}, a Buggy PHP/MySQL Web Application with known vulnerabilities.
This VM has also installed the \texttt{tcpdump}~\cite{b19} v4.9.3 
for capturing all network traffic. 
Client VMs interact with the server and include four web scanners configured to perform SQLi attacks on BWAPP and log their activities: \textit{OWASP ZAP}~\cite{b7} v2.9.0, \textit{Wapiti}~\cite{b8} v3.1.3, \textit{w4af}~\cite{b9} in its pre-alpha release, and \textit{Burp}~\cite{burp} Professional Edition. One client VM hosts the first two scanners, and the other VM hosts the other two. 
We opted only to test one web application, since we only manage the \emph{values} that come in the payloads and the scanners always inject the same inputs, regardless of the target application.

\paragraph{LSH Database Construction}

To construct the LSH database, we used the four scanners and generated four datasets (one for each scanner) of their activity when they carried out SQLi attacks on BWAPP.
Table~\ref{tab:details_lsh} details the used scanners, their entries, and average payload sizes.
During this procedure, a standardised custom scanner identifier was required to normalise the entries (e.g., \texttt{ZAP} to \texttt{scanner}) in the database.
%
We employed the MinHash as the LSH algorithm and created various databases with distinct configurations with $128$ and $256$ permutations and varying threshold values from $0.60$ to $0.95$. 

\paragraph{Query Results Confirmation Process}

To confirm the results of the queries performed on the LSH Database, we developed a two-step method: $i)$ similarity analysis and $ii)$ sanitisation analysis.

For $i)$, the method resorts to an auxiliary file containing the token's keys and their respective inputs from payloads, which was created for that purpose and during the database construction of each scanner, alongside the tokens' generation.
Upon obtaining a query result, the method extracts the tokens linked to the identified keys and computes their \textit{exact Jaccard similarity}. 
If the highest Jaccard similarity between the queried and the returned tokens exceeds the database's threshold, the token should be in the database. The method proceeds with step $ii)$, where if the sanitiser proves the input is an attack (i.e., the sanitiser alters the input), the \tool detection was successful (i.e., a True Positive, TP). Otherwise, the system wrongly identified an attack, causing a False Positive (FP).
If the Jaccard similarity is below the database's threshold or no keys were returned, and the sanitiser (step $ii)$) proves the input is an attack, we consider the system did not identify that attack (i.e., causing a False Negative, FN).

This method also aimed to calculate the \tool's recall (sensitivity), i.e., the percentage of SQLi attacks the system detected as such ($TP/(TP + FN)$), and \tool's precision, i.e.,  the percentage of SQLi attacks
correctly detected by the system ($TP/(TP + FP)$). 

\LONG{
Beyond the experimental testbed specification presented in Section~\ref{sec:implementation}, its setup began with deploying the BWAPP webserver~\cite{b10}, equipped with \texttt{tcpdump}~\cite{b19} v4.9.3 (dependent on \texttt{libpcap} v1.8.1 and OpenSSL v1.1.1d). 
The \texttt{tcpdump} was used for capturing all network traffic. 
Prior to the experiments, the Apache server~\cite{b20} v2.4.38 was prepared by installing the \texttt{mod\_dumpio} add-on and configuring a file to generate necessary logs. 
The server operated on Debian GNU/Linux 10 (buster).

The setup employed four scanners:
\begin{enumerate}
\item \textit{OWASP ZAP}~\cite{b7} v2.9.0, which executed SQLi-focused scans on BWAPP and logged activities. This scanner operated on a system running Kali Linux Rolling, 2020.3 release.
\item \textit{Wapiti} \cite{b8} v3.1.3, employing a similar focus on SQLi attacks. Custom scripts were used for crawling and attacking web pages. It shared the machine and OS specifications with \textit{OWASP ZAP}.
\item \textit{w4af} \cite{b9} in its pre-alpha release, targeting SQLi attacks, required configuration of crawl and audit settings. It ran on a system with Ubuntu 22.04 LTS and Python 3.11.
\item \textit{Burp} \cite{burp} Professional Edition, concentrating on SQLi attacks. Due to its higher computational demands, it utilised the same machine and OS as \textit{w4af}.
\end{enumerate}
} 

\begin{table}[t!]\footnotesize 
    \caption{\small Characterisation of the scanner's datasets}
    \label{tab:details_lsh}
    \vspace{-1mm}
    \centering
    \begin{tabular}{|r|c|c|c|}
    \hline
         \multicolumn{1}{|c|}{\textbf{Dataset}} & \textbf{LSH} &  \textbf{Avg. Payload} &  \textbf{Dataset Size}\\
          \multicolumn{1}{|c|}{\textbf{(Abbrv.)}} &  \textbf{Entries} & \textbf{Size (bytes)} &  \textbf{(bytes)} \\         
         \hline\hline
         ZAP (\colorbox{A5}{\textsc{Za}}) & 8,082 & 87 & 703,134 \\ 
         BURP (\colorbox{A4}{\textsc{Bu}}) & 901 & 118 & 106,318 \\
         W4AF (\colorbox{A3}{\textsc{W$_4$}}) & 122 & 82 & 10,004 \\
         WAPITI (\colorbox{A2}{\textsc{Wa}}) & 72 & 84 & 6,048 \\
         \hline
    \end{tabular}
     \vspace{-3mm}
\end{table}

\subsection{WebFlow Analysis}
\label{ssec:webflow_analysis}

We evaluated WebFlows in two instances: for each scanner, during its database construction stage, and with benign and malicious activity. In both evaluations, WebFlows were driven from traffic captured and processed by the K-Means with the MixMax scaler, which analysed features such as source IP (sIP) and port (sPort), packet number, and duration. 

For the scanners evaluation, the Elbow method indicated an optimal cluster number of $3$ for all scanners, with inertia values close for ZAP ($86$), Wapiti ($83$), w4af ($87$), and Burp ($85$), which suggests the model's effectiveness in identifying varied scanner-generated WebFlows, answering $Q1$ positively.

For the second experiment, we conducted several iterations of benign and malicious activity.
After obtaining $7,728$ WebFlows, we narrowed the number of suspicious ones to $332$. We then applied the Log Extractor module, which resulted in $5,951$ tokens, making it feasible to analyse them using LSH, highlighting the importance of the first phase in identifying web attacks. These tokens are analysed in Section~\ref{ssec:lsh_recall_1}.

\LONG{
\paragraph{Similarity Analysis}

Extensive analysis was conducted on the LSH component. 
Python scripts were developed to process logs from scanners and test them against LSH databases, executing LSH queries and evaluating the results. 
The script assessed the Jaccard similarity between inputs and Minhash keys resulting from the LSH query. 
If the obtained maximum Jaccard similarity is below a threshold or no keys were returned, inputs were sent to a sanitiser for further analysis.
If the sanitiser proves the input is an attack, we consider the LSH could not identify that attack (i.e., causing a False Negative).
Otherwise, the detection was successful (i.e., a True Positive).
This process aimed to calculate the \tool's recall. 
Note the database was not updated during these tests.


}


\subsection{LSH Model Effectiveness}

In this section, we assess the effectiveness of \tool by measuring its sensitivity and precision in detecting SQLi attacks. Sensitivity (or recall) measures how well the model correctly identifies attacks, i.e., TPs and considering FNs, whereas precision measures how many attacks the model correctly detects, i.e., TPs and considering FPs. 

\subsubsection{LSH Model Sensitivity}
\label{ssec:recall}

We performed two types of evaluations to measure the model's sensitivity: $i)$ the model's database includes attacks from a specific scanner to detect other attacks from the same scanner, and $ii)$ the model's database contains attacks from one or more scanners to detect attacks from another scanner.

\noindent\textbf{Attacks from the same Scanner of model's database.}
\label{ssec:lsh_recall}
To accomplish this evaluation, we first observed the system's recall using the ZAP dataset (the biggest dataset of inputs) and varying data ratios to determine the best training/test ratio. Then, we evaluated the other scanner datasets with the identified ratio.
%
The tested ratios and data allocations were:

\begin{itemize}
    \item \textbf{90/10}: $90$\% of the data served as the training set (converted into MinHashes and inserted into the LSH database), and the remaining $10$\% formed the test set (transformed into MinHashes and then queried against the LSH database).
    \item \textbf{80/20}: $80$\% training set and $20$\% test set.
    \item \textbf{70/30}: $70$\% training set and $30$\% test set.
\end{itemize}

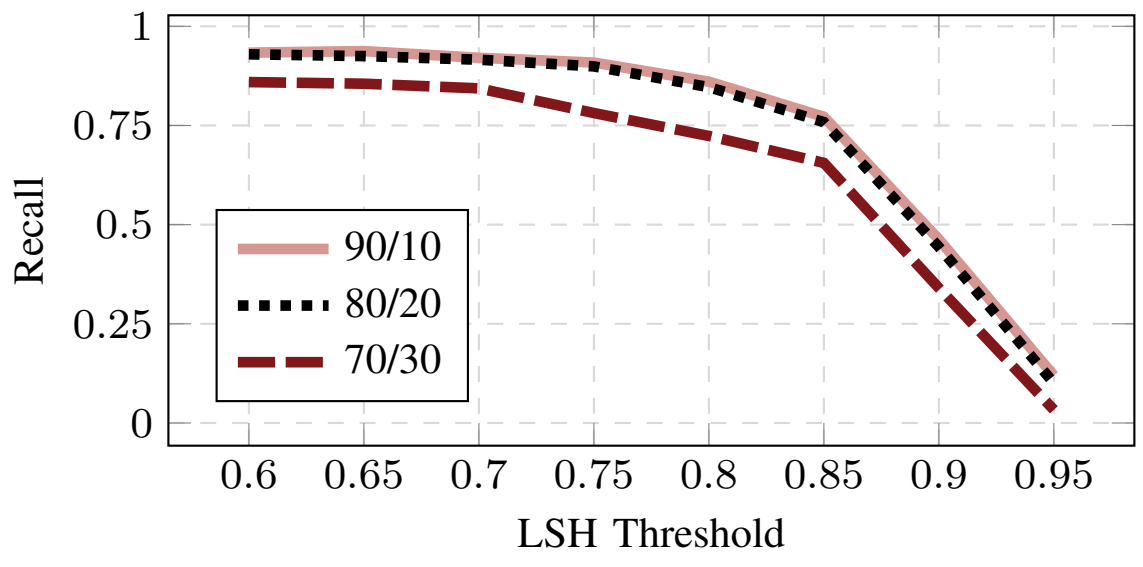
\begin{figure}[b]\footnotesize
    \begin{center}
        \begin{tikzpicture}
            \begin{axis}
                [ width=0.9\columnwidth, 
                  height=0.5\columnwidth,
                  grid=major, 
                  grid style={dashed,gray!30}, 
                  xlabel=LSH Threshold, 
                  ylabel=Recall,
                  ytick distance= 0.25,
                  xtick distance=0.05,
                  legend style={at={(0.18,0.55)},anchor=north}, 
                ]
                \addplot[B3!50, line width=2] table[col sep=comma, x=Threshold, y=Recall ZAP 9010]{graphs/Recall-grafico-comparacao-8020-7030-9010.csv};
                \addlegendentry{\footnotesize 90/10}
                \addplot[black,line width=2,dotted] table[col sep=comma, x=Threshold, y=Recall ZAP 8020]{graphs/Recall-grafico-comparacao-8020-7030-9010.csv};
                \addlegendentry{\footnotesize 80/20}
                \addplot[B2, line width=2,dash pattern=on 7 off 2] table[col sep=comma, x=Threshold, y=Recall ZAP 7030]{graphs/Recall-grafico-comparacao-8020-7030-9010.csv};
                \addlegendentry{\footnotesize 70/30}
            \end{axis}
        
        \end{tikzpicture}
    \caption{\small The recall values of LSH ZAP (\colorbox{A5}{\textsc{Za}}) per threshold with different training/test ratios.}
    \label{fig:graph_ratios}
    \end{center}  
\end{figure}


The outcomes of these tests are depicted in Figure~\ref{fig:graph_ratios}. As expected, the recall values differ across the three training/test ratios at each LSH threshold, reinforcing the idea that the proportion of training to test data 
impacts the model's ability to identify relevant instances correctly. The 90/10 ratio consistently shows higher recall values across all thresholds than the other two ratios. 
This indicates that a more extensive training set ($90$\%) contributes to better model performance in recognising TPs.

We also performed a 10-fold cross-validation to observe the standard deviation ($SD$) in these ratios. The 90/10 ratio resulted in a mean $SD$ of $0.009283$, the 80/20 $0.006936$, and the 70/30 $0.006227$.
Despite the higher recall, the 90/10 ratio also exhibits the highest mean $SD$, implying less consistency in its results. 
This is attributed to the larger test data size in this ratio.
Contrarily, the 70/30 ratio was the worst in both recall and $SD$, and the 80/20 ratio presented the most balanced recall and $SD$, vouching for us to choose it for the next experiment.

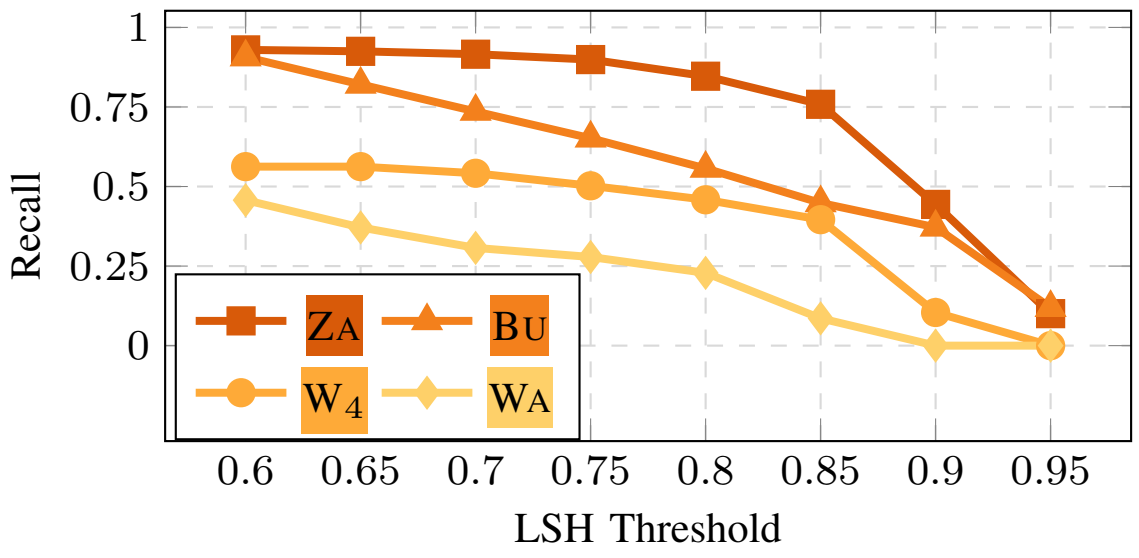
\begin{figure}[t!]\footnotesize
    \begin{center}
        \begin{tikzpicture}
            \begin{axis}
                [
                    width=0.9\columnwidth, 
                    height=0.5\columnwidth,
                  grid=major, 
                  grid style={dashed,gray!30}, 
                  xlabel=LSH Threshold, 
                  ylabel=Recall,
                  ymin=-0.3,
                  legend style={at={(0.22,0.39)},anchor=north,legend columns=2}, 
                  ytick={0,0.25,0.5,0.75,1},
                  xtick distance=0.05,
                ]
                \addplot[A5, mark=cube*, line width=1.5] table[col sep=comma, x=Threshold, y=Recall ZAP]{graphs/Recall-grafico-8020.csv};
                \addlegendentry{\colorbox{A5}{\hspace{-1mm}\textsc{\footnotesize Za}\hspace{-1mm}}}
                \addplot[A4, mark=triangle*, line width=1.5] table[col sep=comma, x=Threshold, y=Recall BURP]{graphs/Recall-grafico-8020.csv};
                \addlegendentry{\colorbox{A4}{\hspace{-1mm}\textsc{\footnotesize Bu}\hspace{-1mm}}}
                \addplot[A3, mark=*, line width=1.5] table[col sep=comma, x=Threshold, y=Recall W4AF]{graphs/Recall-grafico-8020.csv};
                \addlegendentry{\colorbox{A3}{\hspace{-1mm}\textsc{\footnotesize W$_4$}\hspace{-1mm}}}
                \addplot[A2, mark=diamond*, line width=1.5] table[col sep=comma, x=Threshold, y=Recall Wapiti]{graphs/Recall-grafico-8020.csv};
                \addlegendentry{\colorbox{A2}{\hspace{-1mm}\textsc{\footnotesize Wa}\hspace{-1mm}}}
            \end{axis}
        \end{tikzpicture}
    \caption{\small Recall of logs from the different scanners using the selected 80/20 training/testing ratio.}
    \label{fig:recall-comparison}
    \end{center}
     \vspace{-6mm}
\end{figure}

After establishing the 80/20 training/test as the best ratio, we repeat the evaluation for the other scanners.
Figure~\ref{fig:recall-comparison} shows their recalls, indicating that the self-recognition capability among scanners is directly related to the dataset size, i.e., the bigger the dataset, the better the attack identification. 
%
Although ZAP (\textsc{Za}) showed the best self-recognition ability, its recall sharply decreased beyond a threshold of $0.85$, indicating a high degree of similarity among its inputs.
Burp (\textsc{Bu}), ranking second, exhibited a desirable behaviour with a relatively linear decline in recall as the threshold increased, suggesting a balanced and diverse set of inputs.
Although w4af (\textsc{W$_4$}) displayed a pattern similar to \textsc{Za}, implying many analogous inputs, its smaller log size resulted in a lower overall recall.
Finally, Wapiti (\textsc{Wa}) demonstrated the weakest self-identification capability, which can be attributed to its limited sample size in the database.



\noindent\textbf{Attacks from another Scanner of the model's database.}
\label{ssec:lsh_recall_1}
For the evaluation $ii)$, we examined whether an LSH database, constructed using inputs from one or more scanners, can effectively identify attacks from a different scanner.

\begin{table}[t]\footnotesize 
    \caption{\small Recall values with various thresholds for different LSH databases and query datasets.}
    \label{tab:LSH_against_eachother}
    \centering
    \begin{tabular}{|l|l|cccc|}   
    \hline
        \multicolumn{2}{|c|}{\textbf{Dataset}} & \multicolumn{4}{c|}{\textbf{Threshold}} \\
        \hline
        \textbf{Insert} & \textbf{Query} & \textbf{0.6} & \textbf{0.7} & \textbf{0.8} & \textbf{0.9} \\
        \hline
        \hline
        \colorbox{A5}{\textsc{Za}} & \colorbox{A2}{\textsc{Wa}} & 0.5025 & 0.4011 & 0.1302 & 0.0249 \\
        \colorbox{A2}{\textsc{Wa}} & \colorbox{A5}{\textsc{Za}} & 0.2850 & 0.1913 & 0.1192 & 0.0050 \\
        \colorbox{A5}{\textsc{Za}} & \colorbox{A4}{\textsc{Bu}} & 0.4535 & 0.3542 & 0.1184 & 0.0153 \\
        \colorbox{A2}{\textsc{Wa}} & \colorbox{A4}{\textsc{Bu}} & 0.1731 & 0.0560 & 0.0000 & 0.0000 \\
        \colorbox{A5}{\hspace{-1mm}\textsc{Za}}\hspace{-1mm}\colorbox{A2}{\hspace{-1mm}\textsc{Wa}\hspace{-1mm}} & \colorbox{A3}{\textsc{W$_4$}} & 0.5236 & 0.4016 & 0.1390 & 0.0233 \\
        \colorbox{A5}{\hspace{-1mm}\textsc{Za}}\hspace{-1mm}\colorbox{A2}{\hspace{-1mm}\textsc{Wa}}\hspace{-1mm}\colorbox{A4}{\hspace{-1mm}\textsc{Bu}\hspace{-1mm}} & \colorbox{A3}{\textsc{W$_4$}} & 0.5729 & 0.4511 & 0.2032 & 0.0935 \\

       \colorbox{A2}{\hspace{-1mm}\textsc{Wa}}\hspace{-1mm}\colorbox{A4}{\hspace{-1mm}\textsc{Bu}}\hspace{-1mm}\colorbox{A3}{\hspace{-1mm}\textsc{W$_4$}\hspace{-1mm}} &  \colorbox{A5}{\textsc{Za}}$_{332}$ & 0.4876 & 0.4341 & 0.0998 & 0.0513 \\

       \colorbox{A2}{\hspace{-1mm}\textsc{Wa}}\hspace{-1mm}\colorbox{A4}{\hspace{-1mm}\textsc{Bu}}\hspace{-1mm}      \colorbox{A3}{\hspace{-1mm}\textsc{W$_4$}}\hspace{-1mm}\colorbox{A5}{\textsc{Za}}$_{80\%}$  &  \colorbox{A5}{\textsc{Za}}$_{332}$ & 0.9310 & 0.9280 & 0.8513 & 0.4692 \\
        \hline
    \end{tabular}
\end{table}
 
Table~\ref{tab:LSH_against_eachother}, rows 3--8, indicates that the detection rates were consistently higher at lower thresholds ($0.6$). 
Such results can be attributed to the rising demand for similarity in the comparison process, coupled with the fact that not all attack patterns are identical.
%
%
Merging databases from different scanners demonstrated an enhanced ability to detect varied inputs.
However, we noted that the LSH database struggled to recognise inputs other than those originating from its original scanner set, which posed certain limitations.
Moreover, scanners such as Burp, which do not use consistent naming for variable inputs but rather employ random tokens, further complicate the model's effectiveness.
These challenges can potentially be mitigated by regularly updating the LSH database to encompass a broader range of input patterns.

Lastly, we assessed how the model behaves on detecting the $332$ ZAP's WebFlows signalled as malicious (from Section~\ref{ssec:webflow_analysis}), and using for the model's database the other $3$ scanners or these with $80$\% of tokens in the $332$ WebFlows. The last two rows of Table~\ref{tab:LSH_against_eachother} depict the results, which we can weave the same conclusion: recall increases with attacks belonging to the database.
Increasing the similarity threshold decreases the likelihood of a successful recall because it makes harder to match the test tokens with the ones in the database.
The higher the similarity threshold, the more substrings and commands they need to share to create enough common tokens to be considered similar entries.


\LONG{
\begin{figure}[t!]
    \begin{center}
        \begin{tikzpicture}
            \begin{axis}
                [ width=1\columnwidth, 
                  height=0.6\columnwidth,
                  grid=major, 
                  grid style={dashed,gray!30}, 
                  xlabel=LSH Threshold, 
                  ylabel=Recall,
                  ytick distance= 0.25,
                  xtick distance=0.05,
                  legend style={at={(0.2,0.3)},anchor=north}, 
                ]
                \addplot[B3!50, line width=2] table[col sep=comma, x=Threshold, y=Recall]{graphs/WABUW4_vs_ZA_Grafico_recall_threshold.csv};
                \addlegendentry{WaBuW4}
                \addplot[black,line width=2,dotted] table[col sep=comma, x=Threshold, y=Recall]{graphs/WABUW4ZA_vs_ZA_Grafico_recall_threshold.csv};
                \addlegendentry{WaBuW4Za}
            \end{axis}
        
        \end{tikzpicture}
    \caption{\small Caption to be determined}
    \label{fig:graph_332_netflows}
    \end{center}  
\end{figure}

}

\subsubsection{LSH Model Precision}
\label{ssec:precision}


To evaluate the model's precision, we generated benign inputs to measure whether it produced FPs.   
%
The analysis comprised two categories of benign tests:

\begin{enumerate}
    \item[\textbf{T1.}] \textbf{Random Inputs:} Involved $2,021$ strings (inputs) randomly generated\footnote{Generated by \url{https://pynative.com/python-generate-random-string/}} without meta-characters. 
    E.g.:~\texttt{ijarubtd}.
    \item[\textbf{T2.}] \textbf{Scanner-Generated Inputs:} Involved $2,296$ inputs derived from $25$\% of the logs from each of the four scanners. 
    Within this category, three tests were conducted:
    \begin{enumerate}
        \item[\textbf{T2a.}] Trimmed after the first meta-character in the inputs.
        \item[\textbf{T2b.}] Only removed the meta-characters from the inputs.
        \item[\textbf{T2c.}] Removed both meta-characters and SQL commands from the inputs.
    \end{enumerate}
    An example of each T2$_x$, considering the input \texttt{\small bug=41\&form=submit\%22+AND+\%221\%22\%3D\%222} is, respectively, \texttt{\small bug=41\&form=submit}, \\ \texttt{\small bug=41\&form=submit+AND+12}, and \texttt{\small bug=41\&form=submit12}.
    
\end{enumerate}







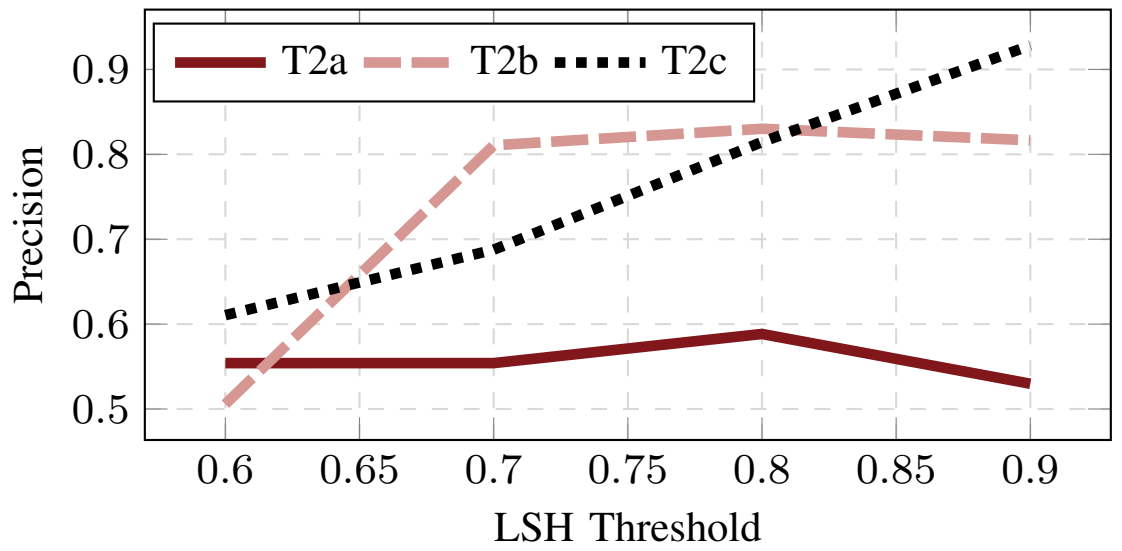
\begin{figure}[t!]\footnotesize
    \begin{center}
        \begin{tikzpicture}
            \begin{axis}
                [
                  width=0.9\columnwidth, 
                  height=0.5\columnwidth,
                  grid=major, 
                  grid style={dashed,gray!30}, 
                  xlabel=LSH Threshold, 
                  ylabel=Precision,
                  ytick={0.5,0.6,0.7,0.8,0.9},
                  legend style={at={(0.32,0.97)},anchor=north,legend columns=3}, 
                  legend cell align={left}, 
                ]
                \addplot[B2, line width=2] table[col sep=comma, x=Threshold, y=Precision2a]{graphs/Precision-grafico.csv};
                \addlegendentry{\footnotesize T2a}
                \addplot[B3!50, line width=2,dash pattern=on 7 off 2] table[col sep=comma, x=Threshold, y=Precision2b]{graphs/Precision-grafico.csv};
                \addlegendentry{\footnotesize T2b}
                \addplot[black, line width=2,dotted] table[col sep=comma, x=Threshold, y=Precision2c]{graphs/Precision-grafico.csv};
                \addlegendentry{\footnotesize T2c}
            \end{axis}
        \end{tikzpicture}
    \caption{\small Precision of the different tests using the LSH ZAP (\colorbox{A5}{\textsc{Za}}).}
    \label{fig:precision-comparison}
    \end{center}
    \vspace{-5mm}
\end{figure}

Outcomes of Test $T1$ were as anticipated, where the LSH configured for ZAP did not register any FPs, aligning with the expectations given the random nature of the inputs. 
In contrast, Test $T2$, as depicted in Figure~\ref{fig:precision-comparison}, presented low precision results. 
This was particularly evident in $T2b$, where the precision was only $50.57$\% at $0.6$ threshold. 
Unlike recall, precision increases with the LSH threshold, improving to $81.65$\% at $0.9$ threshold. 
The most effective modification was observed in $T2c$, where removing meta-characters and SQL commands led to a precision of $61.05$\% at a $0.6$ threshold and an $92.82$\% at a $0.9$ threshold.
The higher the similarity threshold, the better the precision because it reduces the likelihood of dissimilar entries causing False Positives.
These results, albeit low in precision, are derived from extreme case scenarios where samples have prior knowledge of the tokens used to construct the LSH.
It is crucial to note that inputs are only analysed at this stage if their corresponding WebFlow was categorised as malicious.
Hence, an initial stage of FP filtering is inherently present.
This aspect is particularly relevant in the context of potential spoofing attacks, where benign inputs might be intermingled with malicious ones, posing a challenge to \tool.

Overall, from the results we obtained during our experiments (Sections~\ref{ssec:lsh_recall} and~\ref{ssec:precision}), we can state that \tool is able to detect web attacks and confirm their presence in the payloads of the scanners' WebFlows, thus answering to questions $Q2$ and $Q3$.

\subsection{Comparison with ModSecurity}


Lastly, we evaluated \tool with ModSecurity~\cite{modsec}, a widely used Web Application Firewall (WAF) renowned for its ability to defend against web application attacks on HTTP traffic, acting as a DPI system with pattern-based rules.



In assessing the capabilities of the $4$ scanners,
it became apparent that the Burp scanner generated the most complex attacks.
Hence, we chose it to 
launch attacks on the BWAPP application with ModSecurity activated in the Apache server.
Following the attack, we collected and analysed the logs, explicitly identifying the attacks detected by ModSecurity and those it failed to recognise.
This analysis led to an intriguing discovery: ModSecurity did not detect $109$ (out of the $901$) malicious inputs generated by Burp, categorising them as FNs.
This finding resulted in 
a recall rate of $87.9\%$ for ModSecurity. 

To compare \tool with ModSecurity, 
we constructed an LSH database using only inputs from the Burp scanner, deliberately excluding the $109$ undetected inputs, and then we queried the database with these $109$ inputs to evaluate its detection capability. 
At a threshold of $0.6$, \tool correctly recognised $77$ out of $109$ malicious inputs, demonstrating it can detect unseen web attacks and outperforms ModSecurity.

\LONG{
\begin{figure}[t!]\footnotesize
    \begin{center}
        \begin{tikzpicture}
            \begin{axis}
                [
                    width=0.9\columnwidth, 
                    height=0.5\columnwidth,
                  grid=major, 
                  grid style={dashed,gray!30}, 
                  xlabel=LSH Threshold, 
                  ylabel=Recall,
                  ytick distance= 0.1,
                  xtick distance=0.05,
                  legend style={at={(0.2,0.44)},anchor=north}, 
                ]
                \addplot[A6, mark=cube*, line width=1.5] table[col sep=comma, x=Threshold, y=Recall]{graphs/BURP_109FN_ZAP_WAPITI_BURP_Grafico_recall_threshold.csv};
                \addlegendentry{\colorbox{A5}{\hspace{-1mm}\textsc{\footnotesize Za}}\hspace{-1mm}\colorbox{A2}{\hspace{-1mm}\textsc{\footnotesize Wa}}\hspace{-1mm}\colorbox{A4}{\hspace{-1mm}\textsc{\footnotesize Bu}\hspace{-1mm}}}
                \addplot[A4, mark=triangle*, line width=1.5] table[col sep=comma, x=Threshold, y=Recall]{graphs/BURP_contra_109FN_grafico.csv};
                \addlegendentry{\colorbox{A4}{\textsc{\footnotesize Bu}}}
            \end{axis}
        \end{tikzpicture}
    \caption{\small Recall of LSH against the missed inputs of ModSecurity.}
    \label{fig:recall_comparison_109FN}
    \end{center}
\end{figure}
}

\LONG{
\subsection{Cross-Scanner Malicious Input Detection Using Different LSH Databases}
\label{ssec:cross-DB}

Our final experiment examines whether an LSH database, constructed using inputs from a scanner, can effectively identify malicious inputs from a different scanner.
Table~\ref{tab:LSH_against_eachother} indicates that the detection rates were consistently higher at lower thresholds, decreasing notably as the threshold increased.
This pattern can be attributed to the rising demand for similarity in the comparison process, coupled with not all attack patterns being identical.

Although the enhancement in detection was marginal, the capability to update the LSH database suggests that the detection ratios could improve with each update.
A lower number of permutations tends to yield more FPs, whereas a higher threshold reduces the number of keys identified as similar.
This reduction can be beneficial but also introduces a challenge as it requires a larger volume of inputs to undergo sanitisation, a time-consuming process.

Merging databases from different scanners demonstrated an enhanced ability to detect varied inputs.
Additionally, altering the input variable names in tools like ZAP, Wapiti, and w4af to a generic \texttt{fuzzer} label showed slight improvements in detection.
However, we noted that the LSH database struggled to recognise inputs not originating from its initial scanner set, posing certain limitations.
Moreover, scanners such as Burp, which do not use consistent naming for variable inputs but rather employ random tokens, further complicate the effectiveness of the SSE.
These challenges can potentially be mitigated by regularly updating the LSH database to encompass a broader range of input patterns.

Overall, from the results we obtained during our experiments (Sections~\ref{ssec:lsh_recall} to~\ref{ssec:cross-DB}), we can state that \tool is able to detect web attacks and confirm their presence in the payloads of the scanners' WebFlows, thus answering to questions $Q2.$ and $Q3.$.

\begin{table}[t!]\footnotesize  
    \caption{\small Recall values with various thresholds for different LSH databases and query datasets.}
    \label{tab:LSH_against_eachother}
    \centering
    \begin{tabular}{|l|l|cccc|}   
    \hline
        \multicolumn{2}{|c|}{\textbf{Dataset}} & \multicolumn{4}{c|}{\textbf{Threshold}} \\
        \hline
        \textbf{Insert} & \textbf{Query} & \textbf{0.6} & \textbf{0.7} & \textbf{0.8} & \textbf{0.9} \\
        \hline
        \hline
        \colorbox{A5}{\textsc{Za}} & \colorbox{A2}{\textsc{Wa}} & 0.5025 & 0.4011 & 0.1302 & 0.0249 \\
        \colorbox{A2}{\textsc{Wa}} & \colorbox{A5}{\textsc{Za}} & 0.2850 & 0.1913 & 0.1192 & 0.0050 \\
        \colorbox{A5}{\textsc{Za}} & \colorbox{A4}{\textsc{Bu}} & 0.4535 & 0.3542 & 0.1184 & 0.0153 \\
        \colorbox{A2}{\textsc{Wa}} & \colorbox{A4}{\textsc{Bu}} & 0.1731 & 0.0560 & 0.0000 & 0.0000 \\
        \colorbox{A5}{\hspace{-1mm}\textsc{Za}}\hspace{-1mm}\colorbox{A2}{\hspace{-1mm}\textsc{Wa}\hspace{-1mm}} & \colorbox{A3}{\textsc{W$_4$}} & 0.5236 & 0.4016 & 0.1390 & 0.0233 \\
        \colorbox{A5}{\hspace{-1mm}\textsc{Za}}\hspace{-1mm}\colorbox{A2}{\hspace{-1mm}\textsc{Wa}}\hspace{-1mm}\colorbox{A4}{\hspace{-1mm}\textsc{Bu}\hspace{-1mm}} & \colorbox{A3}{\textsc{W$_4$}} & 0.5729 & 0.4511 & 0.2032 & 0.0935 \\
        \hline
    \end{tabular}

\end{table}

} 


\section{Conclusion}
Web applications are central to online services, making them frequent targets of web attacks such as SQL injection.
These attacks are often automated by web scanners that map the application's attack surface, identify vulnerable entry points, and inject malicious input on them.
In this paper, we presented \tool, a NIDS system that detects web application attacks by correlating suspicious WebFlows (NetFlows of HTTP(S) traffic) with endpoint logs.
Specifically, \tool clusters suspicious WebFlows using unsupervised ML and then confirms the presence of attacks in the flagged WebFlows through LSH-based similarity search over tokenised payloads---all without decrypting HTTPS traffic between the client and the server.
The system includes a continuous improvement mechanism that updates the similarity search engine (SSE) as new attacks are observed.
Evaluation results indicate that \tool effectively detects web attacks and outperforms ModSecurity, a widely adopted web application firewall.

\vspace{2mm}
{\small
\noindent\emph{Acknowledgments:}
 \emph{This work was partially supported by P2030 through project I2DT, ref. COMPETE2030-FEDER-00389100, an ITEA4 European project (ref. 22025), and by FCT through the LASIGE Research Unit, ref. UIDB/00408/2025-LASIGE.}
 }

\bibliographystyle{IEEEtran}
\bibliography{main}

\end{document}